\documentclass[manuscript]{aastex}
\usepackage{natbib}
\shorttitle{Chemi-ionization in Solar Photosphere}
\shortauthors{Mihajlov et al.}

\begin{document}

\title{Chemi-ionization in Solar Photosphere:
       Influence on the Hydrogen Atom excited States Population}

\author{Anatolij A. Mihajlov and Ljubinko M. Ignjatovi{\' c}}
\affil{Institute of Physics, University of Belgrade, P.O. Box 57,
11001 Belgrade, Serbia;
\\and
       Isaac Newton Institute of Chile, Yugoslavia Branch, Volgina 7, 11060 Belgrade Serbia}
\email{ljuba@ipb.ac.rs} \email{mihajlov@ipb.ac.rs}

\author{Vladimir A. Sre{\' c}kovi{\' c}}
\affil{Institute of physics, University of Belgrade, P.O. Box 57,
11001 Belgrade, Serbia}

\and

\author{Milan S. Dimitrijevi{\' c}}
\affil{Astronomical Observatory, Volgina 7, 11060 Belgrade,
       Serbia;\\
       and Isaac Newton Institute of Chile, Yugoslavia Branch, Volgina 7, 11060 Belgrade Serbia;\\ and
       Observatoire de Paris, 92195 Meudon Cedex, France}

\begin{abstract}
In this paper, the influence of chemi-ionization processes in $H^*(n
\ge 2) + H(1s)$ collisions, as well as the influence of inverse
chemi-recombination processes on hydrogen atom excited-state
populations in solar photosphere, are compared with the influence of
concurrent electron-atom and electron-ion ionization and
recombination processes. It has been found that the considered
chemi-ionization/recombination processes dominate over the relevant
concurrent processes in almost the whole solar photosphere. Thus, it
is shown that these processes and their importance for the non-LTE
modeling of the solar atmosphere should be investigated further.

\end{abstract}


\section{Introduction}

In order to improve the modeling of the solar photosphere, as well
as to model atmospheres of other similar and cooler stars where the
main constituent is hydrogen too, it is necessary to take into
account the influence of all the relevant collisional processes on
the excited-atom populations in weakly ionized hydrogen plasmas.
This is important for modeling since a strong connection between the
changes in atom excited-state populations and the electron density
exists in weakly ionized plasmas. For example, with an increase of
the electron density, caused by a growth of the excited hydrogen
atom population, the rate of thermalization by electron-atom
collisions in the stellar atmosphere will become higher. A
consequence will be that the radiative source function of the line
center  will be more closely coupled to the Planck function, making
the synthesized spectral lines stronger for a given model structure,
affecting the accuracy of plasma diagnostics and determination of
the atmospheric pressure.

Therefore, in previous papers \citep{mih97, mih03a, mih03b, mih07b},
just a group of chemi-ionization and chemi-recombination atom
collisional processes in weakly ionized layers of stellar
atmospheres (ionization degree less than $10^{-3}$) was studied. In
order to demonstrate the significance of these processes it was
necessary to compare their efficiency, from the aspect of their
influence on the free electron and excited atom populations, with
the efficiency of the known concurrent processes of electron-atom
impact ionization, electron-electron-ion recombination, and
electron-ion photo-recombination. In the helium case, considered in
\citet{mih03a}, it was established that the efficiency of
chemi-ionization and chemi-recombination processes in weakly ionized
layers of the examined DB white dwarf atmospheres was significantly
greater than the efficiency of the relevant electron-atom and
electron-ion processes or at least comparable to them. In the
hydrogen case, considered in \citet{mih97, mih03b, mih07b} in
connection with solar and M red dwarf atmospheres, the relevant
chemi-ionization processes are
%%%%%%%%%%%%%%%%%%%%%%%%%%%%%%%%%%%%%%%%%%%%%%%%%%%%%%%%%%%%%%%%%
\begin{equation}
\label{eq:1a} H^{*}(n) + H(1s)  \Rightarrow   H_{2}^{+} + e,
\end{equation}
%%%%%%%%%%%%%%%%%%%%%%%%%%%%%%%%%%%%%%%%%%%%%%%%%%%%%%%%%%%%%%%%%
\begin{equation}
\label{eq:1b} H^{*}(n) + H(1s)  \Rightarrow H(1s)+H^{+} + e,
\end{equation}
%%%%%%%%%%%%%%%%%%%%%%%%%%%%%%%%%%%%%%%%%%%%%%%%%%%%%%%%%%%%%%%%%
and the corresponding inverse recombination processes are
%%%%%%%%%%%%%%%%%%%%%%%%%%%%%%%%%%%%%%%%%%%%%%%%%%%%%%%%%%%%%%%%%
\begin{equation}\label{eq:2a}
H_{2}^{+}+ e \Rightarrow H^{*}(n) + H(1s),
\end{equation}
%%%%%%%%%%%%%%%%%%%%%%%%%%%%%%%%%%%%%%%%%%%%%%%%%%%%%%%%%%%%%%%%%
\begin{equation}\label{eq:2b}
H(1s)+H^{+}+ e \Rightarrow H^{*}(n) + H(1s),
\end{equation}
%%%%%%%%%%%%%%%%%%%%%%%%%%%%%%%%%%%%%%%%%%%%%%%%%%%%%%%%%%%%%%%%%
where $H^{*}(n)$ is  hydrogen in one of the excited states with the
principal quantum number $n \ge 2$, $H_{2}^{+}$ is the hydrogen
molecular ion in the ground electronic state ($1\Sigma_{g}^{+}$),
and $e$ is a free electron. Consequently, in this case the
efficiency of the chemi-ionization and chemi-recombination processes
has to be compared with the efficiency of the processes
%%%%%%%%%%%%%%%%%%%%%%%%%%%%%%%%%%%%%%%%%%%%%%%%%%%%%%%%%%%%%%%%%
\begin{equation}
\label{eq:3i} H^{*}(n) +  e \Rightarrow   H^{+} + 2 e,
\end{equation}
%%%%%%%%%%%%%%%%%%%%%%%%%%%%%%%%%%%%%%%%%%%%%%%%%%%%%%%%%%%%%%%%%
\begin{equation}
\label{eq:4r} H^{+} + 2 e  \Rightarrow  H^{*}(n) + e,
\end{equation}
%%%%%%%%%%%%%%%%%%%%%%%%%%%%%%%%%%%%%%%%%%%%%%%%%%%%%%%%%%%%%%%%%
\begin{equation}
\label{eq:5r} H^{+}+ e \Rightarrow H^{*}(n) + \varepsilon_{\lambda},
\end{equation}
%%%%%%%%%%%%%%%%%%%%%%%%%%%%%%%%%%%%%%%%%%%%%%%%%%%%%%%%%%%%%%%%%
where $\varepsilon_{\lambda}$ is the energy of a photon with
wavelength $\lambda$.

Let us emphasize the fact that in this paper just the hydrogen case
is at the focus, since our main aim is to draw attention of
astronomers to the processes (\ref{eq:1a}) - (\ref{eq:2b}), and to
show that the importance of these processes for non-LTE modeling of
solar atmosphere should be investigated. For this purpose, it should
be demonstrated that in the solar photosphere the efficiency of
these processes is greater than, or at least comparable to, the
efficiency of processes (\ref{eq:3i}) - (\ref{eq:5r}) within those
ranges of values of $n \ge 2$ and temperature $T$ which are relevant
to the chosen solar atmosphere model. However, by now only for
chemi-recombination processes (\ref{eq:2a}) and (\ref{eq:2b}) it was
qualitatively shown that for $4\le n \le 8$ their efficiency is
comparable with the efficiency of the concurrent processes
(\ref{eq:4r}) and (\ref{eq:5r}) in a part of the solar photosphere
(see \citet{mih97}).

Therefore, the results of new complete calculations are presented
here, which are necessary for achieving the aim of this work. All
the calculations are performed on the basis of the well-known model
C of the solar atmosphere from \citet{ver81}, since it is only for
this model that all the data needed for various calculations are
provided in tabular form. Certainly, we keep in mind also that in
\citet{sti02} this model is cited as practically the single adequate
non-LTE model of the solar atmosphere.

Besides all mentioned, the fact that the processes (\ref{eq:1a}) -
(\ref{eq:2b}) are very important for the solar photosphere is
supported by the results obtained in \citep{mih03b,mih07b}, where
these processes were included ab initio in a non-LTE modeling of an
M red dwarf atmosphere with the effective temperature $T_{eff} =
3800$ K, using PHOENIX code (see \citep{bar98,hau99,sho99}). A fact
was established that including even the
chemi-ionization/recombination only for $4\le n \le 8$, generates
significant changes (by up to 50 percent), at least in the
populations of hydrogen-atom excited states with $2 \le n \le 20$,
and if all these processes (with $n \ge 2$) are included, a
significant change (somewhere up to 2 - 3 times) is also generated
of the free electron density $N_{e}$, and, as one of further
consequences, significant changes in hydrogen line profiles. Keeping
in mind that the compositions of the solar and the considered M red
dwarf's photospheres are practically the same and the values of
hydrogen-atom density, $N_{e}$ and $T$ in these photospheres change
within similar regions \citep{ver81,mih07b}, one can expect that the
influence of processes (\ref{eq:1a}) - (\ref{eq:2b}) on the
hydrogen-atom excited states and free-electron populations in the
solar atmosphere will be at least close to their influence in that
of the M red dwarf, and that these processes will be very important
for weakly ionized layers of the solar atmosphere.

Finally, let us note the fact concerning a group of collision
ion-atom radiative processes. Namely, in several papers
\citep{mih86,mih92b,mih93a,mih94a,mih94b,mih95a,mih07a,ign09} it was
suggested that these processes should be included in the stellar
atmosphere models, and recently it was actually realized in
\citet{fon09}, and \citet{koe10}. Due to a principal similarity
between the mechanisms of processes (\ref{eq:1a}) - (\ref{eq:2b})
and these radiative processes, one can hope that the
chemi-ionization/recombination processes will be also included in
the stellar atmosphere models.

In this paper all the needed theoretical data, about the
chemi-ionization and chemi-recombination processes (\ref{eq:1a}) -
(\ref{eq:2b}) are given in Section~\ref{sec:theory}, while in
Section~\ref{sec:RandD} the obtained results are presented (in
figures) demonstrating the significance of the considered processes.
Apart from that, all the data used for the calculation of the rate
coefficients of processes (\ref{eq:1a}) - (\ref{eq:2b}) are given
here in three tables.

\section{Theory}
\label{sec:theory}

\subsection{The chemi-ionization processes: $n\ge 5$}
\label{sec:n5ci}

Here we will consider processes (\ref{eq:1a}) - (\ref{eq:2b}) within
the regions $n\ge 5$ and $2\le n \le 4$ separately. Within the first
region we will determine the rate coefficients for the
chemi-ionization processes (\ref{eq:1a}) and (\ref{eq:1b}) directly,
using the principle of thermodynamical balance for determination of
the rate coefficients for the inverse chemi-recombination processes
(\ref{eq:2a}) and (\ref{eq:2b}).

The parameters which are needed in further considerations are the
following: $r_{n}\sim n^{2}$ - the characteristic radius of Rydberg
atom $H^{*}(n)$; $R$ - the inter-nuclear distance in the collision
system $H^{*}(n) + H(1s)$; $U_{1}(R)$ and $U_{2}(R)$ - the adiabatic
potential energies of the ground and the first exited electronic
states of molecular ion $H_{2}^{+}$.

In accordance with the previous papers
\citep{mih97,mih03a,mih03b,mih07b} we will treat processes
(\ref{eq:1a}) and (\ref{eq:1b}) with $n\ge 5$ on the basis of dipole
resonant mechanism, which was introduced in the considerations of
\citet{smi71} for inelastic processes in thermal $[H^{*}(n) +
H(1s)]$-collisions. This means that such processes are considered as
a result of resonant energy conversion within the electronic
component of the collisional system $H^{*}(n) + H(1s)$, which is
realized inside the region
%%%%%%%%%%%%%%%%%%%%%%%%%%%%%%%%%%%%%%%%%%%%%%%%%%%%%%%%%%%%%%%%%%%
\begin{equation}\label{eq:qmol}
R << r_{n},
\end{equation}
%%%%%%%%%%%%%%%%%%%%%%%%%%%%%%%%%%%%%%%%%%%%%%%%%%%%%%%%%%%%%%%%%%%
where the system $H^{*}(n) + H(1s)$ can be presented as $[H^{+}+
H(1s)]+ e_{n}$, and which is caused by the dipole part of the
interaction of the outer electron $e_{n}$ with the subsystem $[H^{+}
+ H(1s)]$. Already in \citet{dev78} just chemi-ionization processes
in atom-Rydberg-atom collisions (the case of alkali metal atoms)
were described on the basis of the same mechanism. After that the
methods based on the dipole resonant mechanism have been used in
practice for investigation of chemi-ionization processes until now
(see for example \citet{bez03} and \citet{ign05}). Application of
this mechanism is particularly successful within the so-called decay
approximation which was examined in \citet{jan80} and immediately
demonstrated to be suitable for any inelastic processes in slow
$[H^{*}(n) + H(1s)]$-collisions. Let us note that in the previous
papers \citep{mih03a,mih03b,mih07b} a method from \citet{mih92b} and
\citet{mih96} was used, which is based on this approximation.

Only the decay approximation will be used here for processes
(\ref{eq:1a}) and (\ref{eq:1b}) with $n\ge 5$. First, it is assumed
that in the region, Equation.~(\ref{eq:qmol}), the electronic state
of the subsystem  $[H^{+} + H(1s)]$ can be approximated well by one
of the two adiabatic electronic states of the molecular ion
$H_{2}^{+}$: the ground one $|1\Sigma_{g}^{+}\vec{r}_{mi},R>$ and
the first excited $|1\Sigma_{u}^{+}\vec{r}_{mi},R>$, and the state
of the outer electron $e_{n}$ - by one of the hydrogen Rydberg
states $|n,l,m,\vec{r}>$. Then, it is assumed that, in the case of
chemi-ionization processes (\ref{eq:1a}) and (\ref{eq:1b}), we can
consider that the subsystem $[H^{+} + H(1s)]$ is in the first
excited electronic state $|1\Sigma_{u}^{+}\vec{r}_{mi},R>$ with a
probability of 1/2, which means that we can describe the relative
inter-nuclear motion as going on in the reflective potential
$U_{2}(R)$. Finally, it means that we can expect (as a result of the
mentioned electron-dipole interaction) a decay of the initial
electronic state $|n,l,m,\vec{r};1\Sigma_{u}^{+}\vec{r}_{mi},R> =
|n,l,m,\vec{r}> |1\Sigma_{u}^{+} \vec{r}_{mi},R>$ of the system
$[H^{+}+ H(1s)] + e_{n}$, with a transition to the final state
$|\epsilon,l',m',\vec{r}; 1\Sigma_{g}^{+}\vec{r}_{mi},R> =
|\epsilon,l',m',\vec{r}>|1\Sigma_{g}^{+} \vec{r}_{mi},R>$ in a
narrow neighborhood of the resonant point $R = R_{n;\epsilon}$ which
is the root of the equation
%%%%%%%%%%%%%%%%%%%%%%%%%%%%%%%%%%%%%%%%%%%%%%%%%%%%%%%%%%%%%%%%%%%
\begin{equation} \label{eq:zona}
U_{12}(R)\equiv U_{2}(R)- U_{1}(R) = \epsilon -\epsilon_{n}.
\end{equation}
%%%%%%%%%%%%%%%%%%%%%%%%%%%%%%%%%%%%%%%%%%%%%%%%%%%%%%%%%%%%%%%%%%%
Here $\epsilon_{n}\cong -I/n^{2}$ and  $\epsilon$ are the energies
of the initial (bound) and final (free) states of the outer
electron, and $I$ - the ionization potential of the ground-state
hydrogen atom. Here it is important that we have almost resonant
simultaneous transitions: of the subsystem $[H^{+}+ H(1s)]$ to the
ground electronic state $|1\Sigma_{g}^{+} \vec{r}_{mi}, R>$, and of
the outer electron to one of the free states $|\epsilon,l', m',
\vec{r}>$.

In accordance with \citep{jan80,mih92b,mih96} we will characterize
processes (\ref{eq:1a}) and (\ref{eq:1b}) by quantities $W_{n}(R)$,
$P_{ci}^{(a)}(n;\rho,E)$, and $P_{ci}^{(b)}(n;\rho,E)$. The first
quantity has the meaning of mean decay velocity of the considered
system's initial state and is given by relations
%%%%%%%%%%%%%%%%%%%%%%%%%%%%%%%%%%%%%%%%%%%%%%%%%%%%%%%%%%%%%%%%%%%%
\begin{equation} \label{eq:W}
W_{n}(R) = \frac{1}{n^{2}}\cdot\sum\limits_{l,m} \frac{2\pi}{\hbar}
\cdot |<1u; n,l,m,|e^{2}\cdot\frac{\vec{r}\cdot\vec{r}_{mi}}{r^{3}}|
\epsilon_{k}, l',m'; 1g^{+}>|^{2} \cdot g(\epsilon),
\end{equation}
%%%%%%%%%%%%%%%%%%%%%%%%%%%%%%%%%%%%%%%%%%%%%%%%%%%%%%%%%%%%%%%%%%%%
where $e$ is the electron charge, $|1g,\epsilon,l',m'> =
|\epsilon,l',m', \vec{r}>|1\Sigma_{g}^{+} \vec{r}_{mi},R>$,
$|1u,n,l,m> = |n,l,m,\vec{r}> |1\Sigma_{u}^{+} \vec{r}_{mi},R>$, and
$g(\epsilon)$ is the density of the free single-electron states in
the energy space. Following \citet{jan80} and \citet{ign05} we will
transform Equation~(\ref{eq:W}) to the simple form
%%%%%%%%%%%%%%%%%%%%%%%%%%%%%%%%%%%%%%%%%%%%%%%%%%%%%%%%%%%%%%%%%%%%
\begin{equation}\label{eq:Wn}
W_{n}(R) = \frac{4}{3\sqrt{3}n^{5}} D_{12}^{2}(R) G_{nk}, \quad
D_{12}(R) = |<1\Sigma_{u}^{+}\vec{r}_{mi},R|r_{mi;R}|1\Sigma_{g}^{+}
\vec{r}_{mi},R>|,
\end{equation}
%%%%%%%%%%%%%%%%%%%%%%%%%%%%%%%%%%%%%%%%%%%%%%%%%%%%%%%%%%%%%%%%%%%%
where $r_{mi;R}$ is the projection of $\vec{r}_{mi}$ on the
inter-nuclear axis, $G_{nk}\equiv
\sigma_{ph}(n,k)/\sigma_{ph}^{Kr}(n,k)$ is the generalized Gaunt
factor, defined in \citet{joh72}, $\sigma_{ph}(n,k)$ is the mean
photo-ionization cross section of the atom $H^{*}(n)$ for the given
$\epsilon$, and $\sigma_{ph}^{Kr}(n,k)$ is the same photo-ionization
cross section, but in Kramers's approximation \citep{kra23,sob79}.

The quantities $P_{ci}^{(a)}(n;\rho,E)$ and $P_{ci}^{(b)}(n;\rho,E)$
are the probabilities of the realization of the chemi-ionization
processes (\ref{eq:1a}) and (\ref{eq:1b}) respectively, for given
$\rho$ and $E$. In the previous papers
\citep{mih92b,mih96,mih03a,mih03b,mih07b}, the influence of the
initial state's decay (during the collision) on its amplitude was
neglected in order to simplify the used procedure. However, it
generates errors which are larger then $10$ percent for $n\le 8$.
Consequently, in this work the said influence is taken into account
and a procedure similar to the ones from \citet{jan80} and
\citet{ign05} is used. Therefore we will present here only the final
expressions for the ionization probabilities
$P_{ci}^{(a,b)}(n;\rho,E)$, the partial cross-sections
$\sigma_{ci}^{(a,b)}(n;E)$ and the corresponding partial rate
coefficients $K_{ci}^{(a,b)}(n;T)$, where $T$ is the temperature of
the considered plasma, using additional parameters $R_{n}$, $R_{0}$
and $R_{E}$, which are the roots of equations
%%%%%%%%%%%%%%%%%%%%%%%%%%%%%%%%%%%%%%%%%%%%%%%%%%%%%%%%%%%%%%%%%%%%
\begin{equation} \label{eq:RRR}
U_{12}(R)= |\epsilon_{n}|, \qquad  U_{2}(R)=E, \qquad  U_{12}(R)=E,
\end{equation}
%%%%%%%%%%%%%%%%%%%%%%%%%%%%%%%%%%%%%%%%%%%%%%%%%%%%%%%%%%%%%%%%%%%%
respectively.

In the case when only one of the processes (\ref{eq:1a}) and
\ref{eq:1b}) is realized, the ionization probabilities are obtained
in the form
%%%%%%%%%%%%%%%%%%%%%%%%%%%%%%%%%%%%%%%%%%%%%%%%%%%%%%%%%%%%%%%%
\begin{equation}\label{eq:Pfin1}
P_{ci}^{(a,b)}(n,\rho,E)= \frac{1}{2} \cdot \left(1-e^{-2
\int\limits_{R_{0}}^{R_{n}} \frac{W_{n}(R)dR}{v_{rad}}}\right),
\end{equation}
%%%%%%%%%%%%%%%%%%%%%%%%%%%%%%%%%%%%%%%%%%%%%%%%%%%%%%%%%%%%%%%%
and in the case of realization of both processes we have it that
%%%%%%%%%%%%%%%%%%%%%%%%%%%%%%%%%%%%%%%%%%%%%%%%%%%%%%%%%%%%%%%%
\begin{equation}\label{eq:Pfin2}
P_{ci}^{(a)}(n,\rho,E)= \frac{1}{2} \cdot
\left(1-e^{-2\int\limits_{R_{0}}^{R_{E}}
\frac{W_{n}(R)dR}{v_{rad}}}\right) e^{-\int\limits_{R_{E}}^{R_{n}}
\frac{W_{n}(R)dR}{v_{rad}}},
\end{equation}
%%%%%%%%%%%%%%%%%%%%%%%%%%%%%%%%%%%%%%%%%%%%%%%%%%%%%%%%%%%%%%%%
\begin{equation}\label{eq:Pfin3}
P_{ci}^{(b)}(n,\rho,E)= \frac{1}{2} \cdot
\left(1-e^{-\int\limits_{R_{E}}^{R_{n}}
\frac{W_{n}(R)dR}{v_{rad}}}\right)
\left(1+e^{-2\int\limits_{R_{0}}^{R_{n}} \frac{W_{n}(R)dR}{v_{rad}}
}\right),
\end{equation}
%%%%%%%%%%%%%%%%%%%%%%%%%%%%%%%%%%%%%%%%%%%%%%%%%%%%%%%%%%%%%%%%
where $\rho$ and $E = m_{red}v^{2}/2$ are the impact parameter and
the atom-Rydberg-atom impact energy, respectively ($m_{red}$ being
the reduced mass of the collision system)

$v_{rad}= v_{rad}(\rho,E;R)$ is the radial inter-nuclear velocity,
which is given by
%%%%%%%%%%%%%%%%%%%%%%%%%%%%%%%%%%%%%%%%%%%%%%%%%%%%%%%%%%%%%%%%
\begin{equation}\label{eq:vrad}
v_{rad}(\rho,E;R)=\sqrt{\displaystyle
\frac{2}{m_{red}}\left[E-U_{2}(R)-\frac{E\rho^{2}}{R^{2}}\right]}.
\end{equation}
%%%%%%%%%%%%%%%%%%%%%%%%%%%%%%%%%%%%%%%%%%%%%%%%%%%%%%%%%%%%%%%%
Then, from Equations.~(\ref{eq:Pfin1}) - (\ref{eq:vrad}) the partial
cross sections $\sigma_{ci}^{(a,b)}(n;E)$ are determined, namely,
%%%%%%%%%%%%%%%%%%%%%%%%%%%%%%%%%%%%%%%%%%%%%%%%%%%%%%%%%%%%%%%%
\begin{equation}\label{eq:sigma}
\sigma_{ci}^{(a,b)}(n,E) = 2 \pi
\int\limits_{0}^{\rho_{max}^{(a,b)}(E)} {P_{ci}^{(a,b)}(n,\rho,E)
\rho} d\rho,
\end{equation}
%%%%%%%%%%%%%%%%%%%%%%%%%%%%%%%%%%%%%%%%%%%%%%%%%%%%%%%%%%%%%%%%
where $\rho^{max}_{(1a,b)}(E)$ is the upper limit of values $\rho$,
at which the corresponding region $R$ is reached for a given $E$.

After that, the partial rate coefficients for the chemi-ionization
processes (\ref{eq:1a}) and (\ref{eq:1b}) with $n\ge 5$ are
determined by expressions
%%%%%%%%%%%%%%%%%%%%%%%%%%%%%%%%%%%%%%%%%%%%%%%%%%%%%%%%%%%%%%%%
\begin{equation}\label{eq:Kn}
K_{ci}^{(a,b)}(n,T) = \int\limits_{E_{min}^{(a,b)}(n)}^{E_{max}} {v
\sigma_{ci}^{(a,b)}(n,E) f(v;T)} dv,
\end{equation}
%%%%%%%%%%%%%%%%%%%%%%%%%%%%%%%%%%%%%%%%%%%%%%%%%%%%%%%%%%%%%%%%
where $\sigma_{ci}^{(a,b)}(n,E)$ is defined by Equation
(\ref{eq:sigma}), $v$ is the atom-Rydberg-atom impact velocity,
$f(v;T)$ is the velocity distribution function for the given
temperature $T$, and: $E_{min}^{(a,b)}(n) = 0$ if $U_{2}(R_{n}) \le
0$; $E_{min}^{(a,b)} = U_{2}(R_{n})$ if $U_{2}(R_{n}) > 0$;
$E_{max}^{(a)} = U_{2}(R_{0;1}) = U_{2}(R_{0;1})$, where $R_{0;1}$
is such a point that $U_{1}(R_{0;1})=0$; $E_{max}^{(b)}=\infty$.

Finally, using partial rate coefficients $K_{ci}^{(a,b)}(n,T)$, we
will determine the total one, namely,
%%%%%%%%%%%%%%%%%%%%%%%%%%%%%%%%%%%%%%%%%%%%%%%%%%%%%%%%%%%%%%%%
\begin{equation}\label{eq:Kzbir}
K_{ci}(n,T)=K_{ci}^{(a)}(n,T)+K_{ci}^{(b)}(n,T),
\end{equation}
%%%%%%%%%%%%%%%%%%%%%%%%%%%%%%%%%%%%%%%%%%%%%%%%%%%%%%%%%%%%%%%%
which characterizes the efficiency of the chemi-ionization processes
(\ref{eq:1a}) and (\ref{eq:1b}) together.

\subsection{The chemi-recombination processes: $n\ge 5$}
\label{sec:n5cr} Under the conditions which exist in the solar
atmosphere, we can determine the chemi-recombination rate
coefficients (as functions of $T$) from the principle of
thermodynamical balance for processes (\ref{eq:1a},\ref{eq:1b}) and
(\ref{eq:2a},\ref{eq:2b}), namely,
%%%%%%%%%%%%%%%%%%%%%%%%%%%%%%%%%%%%%%%%%%%%%%%%%%%%%%%%%%%%%%%%
\begin{equation}\label{eq:Kra}
K_{ci}^{(a)}(n,T)\cdot N_{n} N_{1} = K_{dr}(n,T)\cdot N_{mi}^{(eq)}
N_{e} \equiv K_{cr}^{(a)}(n,T)\cdot N_{1}N_{ai}N_{e}
\end{equation}
%%%%%%%%%%%%%%%%%%%%%%%%%%%%%%%%%%%%%%%%%%%%%%%%%%%%%%%%%%%%%%%%
\begin{equation}\label{eq:Krb}
K_{ci}^{(b)}\cdot N_{n} N_{1} = K_{cr}^{(b)}(n,T)\cdot
N_{1}N_{ai}N_{e},
\end{equation}
%%%%%%%%%%%%%%%%%%%%%%%%%%%%%%%%%%%%%%%%%%%%%%%%%%%%%%%%%%%%%%%%
where the chemi-ionization rate coefficient $K_{cr}^{(a)}(n,T)$ is
expressed through the dissociative recombination rate coefficient
$K_{dr}(n,T)$ by relation
%%%%%%%%%%%%%%%%%%%%%%%%%%%%%%%%%%%%%%%%%%%%%%%%%%%%%%%%%%%%%%%%
\begin{equation}\label{eq:Kdra}
K_{cr}^{(a)}(n,T) \equiv K_{dr}(n,T)\cdot \chi^{-1}(T), \qquad
\chi(T) = \left(\frac{N_{ai}N_{1}}{N_{mi}^{(eq)}}\right),
\end{equation}
%%%%%%%%%%%%%%%%%%%%%%%%%%%%%%%%%%%%%%%%%%%%%%%%%%%%%%%%%%%%%%%%
where $N_{1}^{}$ and $N_{n}^{}$ denote the densities of ground- and
excited-state hydrogen atoms respectively, while $N_{ai}$ and
$N_{mi}^{(eq)}$ are the densities of atomic ions $H^{+}$ and
molecular ions $H_{2}^{+}$ respectively. The index $"eq"$ denotes
that molecular ion density corresponds to thermodynamical
equilibrium condition for given $T$. Factor $\chi(T)$ can be
determined as in \citet{mih07a} in connection with the contribution
of $H^{+} + H(1s)$ radiative collision processes to the solar
atmosphere's opacity in UV and VUV region.

Taking quantity $K_{cr}^{(a)}(n,T)$ as the rate coefficient for
process (\ref{eq:2a}), we can characterize both chemi-recombination
processes (\ref{eq:2a}) and (\ref{eq:2b}) in a similar way. Namely,
in accordance with Equations~(\ref{eq:Kra}) and (\ref{eq:Krb}), rate
coefficients $K_{cr}^{(a)}(n,T)$ and $K_{cr}^{(b)}(n,T)$ are given
by relations
%%%%%%%%%%%%%%%%%%%%%%%%%%%%%%%%%%%%%%%%%%%%%%%%%%%%%%%%%%%%%%%%
\begin{equation}\label{eq:DB}
K_{cr}^{(a,b)}(n,T) = K_{ci}^{(a,b)}(n,T)\cdot S_{n}^{-1}(T), \qquad
S_{n}(T) \equiv \frac{N_{i}N_{e}}{N_{n}} = \frac{1}{n^{2}}\cdot
\frac{mk_{B}T}{2\pi \hbar^{2}} \cdot \exp(-\frac{I_{n}}{k_{B}T}),
\end{equation}
%%%%%%%%%%%%%%%%%%%%%%%%%%%%%%%%%%%%%%%%%%%%%%%%%%%%%%%%%%%%%%%%
where $m$ is the electron mass and $k_{B}$ is the Boltzmann
constant. Consequently, using such partial rate coefficients, we can
introduce here the total one, i.e.,
%%%%%%%%%%%%%%%%%%%%%%%%%%%%%%%%%%%%%%%%%%%%%%%%%%%%%%%%%%%%%%%%
\begin{equation}\label{eq:Kr0}
K_{cr}(n,T)=K_{cr}^{(a)}(n,T)+K_{cr}^{(b)}(n,T),
\end{equation}
%%%%%%%%%%%%%%%%%%%%%%%%%%%%%%%%%%%%%%%%%%%%%%%%%%%%%%%%%%%%%%%%
which characterizes the efficiency of processes \ref{eq:2a}) and
(\ref{eq:2b}) together for $n\ge 5$.

\subsection{The chemi-ionization/recombination processes:
$2\le n\le 4$} \label{sec:n2cir}

The reason why the regions $n\ge 5$ and $2\le n \le 4$ are being
considered separately is the behavior of the adiabatic potential
curves of atom-atom systems $H^{*}(n) + H(1s)$. Namely, in the first
region the atom-atom curves lie above the adiabatic curve of the
ion-ion system $H^{+} + H^{-}(1s^{2})$ for any $R$, and the dipole
resonant mechanism can be applied for $n\ge 5$ without any
exceptions, while in the other region there are points where the
atom-atom curves cross the ion-ion one and application of this
mechanism generates some errors (see \cite{jan79}). Since the
corresponding cross-points for $n\le4$ are so far from the point
$R=0$ that their existence could be neglected for $n=4$ and $3$, and
with some caution even for $n=2$, the dipole resonant mechanism was
applicable, for example, in \citet{mih92b} and \citet{mih96} for
$n=4$ and in \citet{zhd76} for $n=3$. However, now we can determine
the values of rate coefficients $K_{cr}^{(a)}(n,T)$ of dissociative
recombination process (\ref{eq:2a}) for $n=2$, $3$, and $4$ using
the results deduced from the experimental data of \citet{jon77},
presented in \citet{jan87}.

Due to this fact and the mentioned errors, we use here
semi-empirical rate coefficients $K_{cr}^{(a)}(n=3,T)$ and
$K_{cr}^{(a)}(n=4,T)$, which are obtained from \citet{jan87}, for
the dominant processes of the dissociative recombination, i.e., for
process (\ref{eq:2a}) with $n=3$ and $4$. The corresponding
chemi-ionization rate coefficients $K_{ci}^{(a)}(n=3,T)$ and
$K_{ci}^{(a)}(n=4,T)$ are obtained then from the principle of
thermodynamical balance, as it has been described above. For
relatively minor chemi-ionization/recombination processes, i.e. for
processes (\ref{eq:1a}) and (\ref{eq:2a}) with $n=2$, we use here
rate coefficients $K_{ci}^{(a)}(n=2,T)$ and $K_{cr}^{(a)}(n=2,T)$,
which are $10$ - $30$ percent greater than the corresponding
coefficients obtained using the data from \citet{jan87}, in
accordance with the calculated results from \citet{urb91} and
\citet{raw93}. It gives a possibility to compensate the decrease of
rate coefficients $K_{ci}^{(a)}(n\ge 5,T)$ and $K_{cr}^{(a)}(n\ge
5,T)$ in comparison with the corresponding ones obtained using
\citet{jan87}, due to the fact that here, unlike \citet{jan87}, the
decay of the considered system's initial electronic state has been
taken into account. For other chemi-ionization and recombination
processes (\ref{eq:1b}) and (\ref{eq:2b}) with $2\le n \le 4$, whose
contribution could really be neglected, the corresponding rate
coefficients will be determined (in accordance with what was said
above) by extrapolation of those from the region $n\ge 5$. Finally,
let us note that in further considerations for chemi-ionization and
recombination processes (\ref{eq:1a}) - (\ref{eq:2b}) with $n < 5$
we will use also total rate coefficients, which are given by the
same expressions (\ref{eq:Kzbir}) and (\ref{eq:Kr0}), but for $2\le
n \le 4$.

\section{Results and discussion}
\label{sec:RandD}

\subsection{The Considered Model of the Solar Photosphere}

\begin{figure}
\centerline{\includegraphics[width=\columnwidth,
height=0.75\columnwidth]{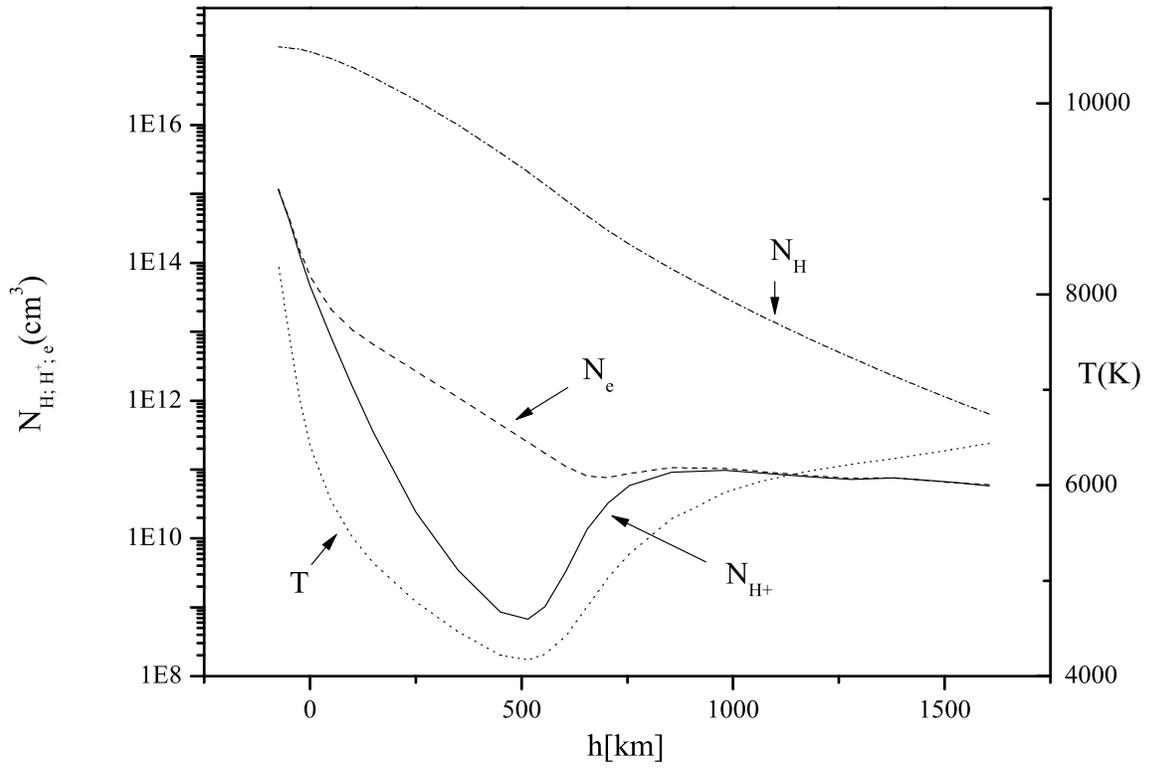}} \caption{Basic plasma
parameters, for the solar model of \citet{ver81}, as a function of
height h.} \label{fig:plasma_parameters}
\end{figure}

\begin{figure}
\centerline{\includegraphics[width=\columnwidth,
height=0.75\columnwidth]{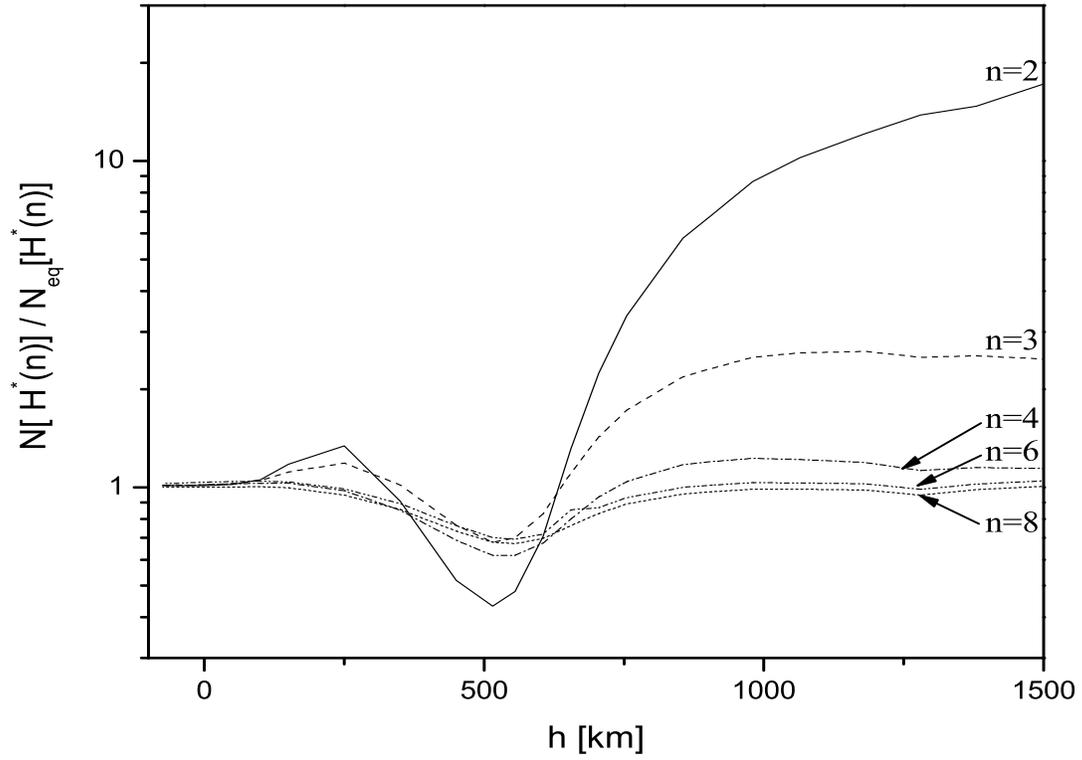}} \caption{Parameter
$\eta(n)=N(H^{*})(n)/N^{(eq)}(H^{*})(n)$, as a function of height h.
The index "eq" denotes that excited atom densities correspond to
thermodynamical equilibrium conditions for given $T$.}
\label{fig:eta}
\end{figure}

In accordance with the aim of this work we consider here model C of
solar atmosphere from \citet{ver81}. Namely, this is a non-LTE model
which is still actual (see \citep{sti02}), and it is only for this
model that all the quantities necessary for our calculations are
available in tabular form as functions of height ($h$) in Solar
photosphere. In Figure \ref{fig:plasma_parameters}, basic plasma
parameters for this model are shown. In Figure \ref{fig:eta},
deviations of non-LTE populations of excited hydrogen atom states
with $2\le n \le 8$ in solar photosphere within the C model of
\citet{ver81} are illustrated. One can see that these deviations are
particularly pronounced for $n$ = 2. Around $h$ = 500 km
$N\left(H^{*}(n=2)\right)$ is one-half of the corresponding
equilibrium density and for $h$ larger than 1000 km it is around ten
times greater. These deviations rapidly decrease with an increase of
$n$. However, even for $n$=8 this deviation is around 40 percent
around $h$ = 500 km, illustrating the importance of taking into
account the considered processes ab initio in the modeling of solar
atmosphere.

\subsection{The calculated chemi-ionization/recombination rate coefficients}

The values of the total chemi-ionization and recombination rate
coefficients $K_{ci}(n,T)$ and $K_{cr}(n,T)$, obtained in the
described way, are presented in Tables~\ref{tab:Kci} and
\ref{tab:Kcr} respectively. These tables cover the regions $2\le
n\le 8$ and $4000 K\le T\le 10 000 K$ which are relevant for solar
photosphere considered on the basis of C model from \citep{ver81}.

Relative contribution of partial chemi-ionization and recombination
processes for given $n$ and $T$ characterizes corresponding branch
coefficients $X_{ci}^{(a,b)}(n,T)$, namely
%%%%%%%%%%%%%%%%%%%%%%%%%%%%%%%%%%%%%%%%%%%%%%%%%%%%%%%%%%%%%%%%
\begin{equation}\label{eq:X}
X_{ci}^{(a,b)}(n,T)=\frac{K_{ci}^{(a,b)}(n,T)}{K_{ci}(n,T)}, \qquad
X_{cr}^{(a,b)}(n,T) = \frac{K_{cr}^{(a,b)}(n,T)}{K_{cr}(n,T)}.
\end{equation}
%%%%%%%%%%%%%%%%%%%%%%%%%%%%%%%%%%%%%%%%%%%%%%%%%%%%%%%%%%%%%%%%
Since $X_{ci, cr}^{(b)}(n,T) = 1 -X_{ci, cr}^{(a)}(n,T)$ and
$X_{ci}^{(a,b)}(n,T) = X_{cr}^{(a,b)}(n,T) \equiv X^{(a,b)}(n,T)$,
it is enough to present only the values of one of the coefficients
$X^{(a,b)}(n,T)$. Here, the values of the coefficient
$X^{(a)}(n,T)$, which directly describe relative contributions of
the associative ionization and dissociative recombination processes
(\ref{eq:1a}) and (\ref{eq:2a}), are presented in Table~\ref{tab:X}.

\subsection{Comparison of fluxes of the considered processes}

Let $I_{ci}(n,T)$, $I_{cr}(n,T)$ be the total chemi-ionization and
chemi-recombination fluxes caused by the processes
(\ref{eq:1a},\ref{eq:1b}) and (\ref{eq:2a},\ref{eq:2b}), i.e.,
%%%%%%%%%%%%%%%%%%%%%%%%%%%%%%%%%%%%%%%%%%%%%%%%%%%%%%%%%%%%%%%%
\begin{equation}\label{eq:I12}
I_{ci}(n,T) = K_{ci}(n,T)\cdot N_{n}N_{1}, \qquad I_{cr}(n,T) =
K_{cr}(n,T) \cdot N_{1}N_{i}N_{e},
\end{equation}
%%%%%%%%%%%%%%%%%%%%%%%%%%%%%%%%%%%%%%%%%%%%%%%%%%%%%%%%%%%%%%%%
and $I_{i;ea}(n,T)$, $I_{r;eei}(n,T)$ and $I_{r;ph}(n,T)$ be the
fluxes caused by ionization and recombination processes
(\ref{eq:3i}), (\ref{eq:4r}) and (\ref{eq:5r}), i.e.
%%%%%%%%%%%%%%%%%%%%%%%%%%%%%%%%%%%%%%%%%%%%%%%%%%%%%%%%%%%%%%%%
\begin{equation}\label{eq:I345}
I_{i;ea}(n,T) = K_{ea}(n,T)\cdot N_{n}N_{e}, \quad I_{r;eei}(n,T) =
K_{eei}(n,T) \cdot N_{i}N_{e}N_{e},  \quad I_{r;ph}(n,T) =
K_{ph}(n,T)\cdot N_{i}N_{e},
\end{equation}
%%%%%%%%%%%%%%%%%%%%%%%%%%%%%%%%%%%%%%%%%%%%%%%%%%%%%%%%%%%%%%%%
where $N_{1}$, $N_{n}$, $N_{i}$, and $N_{e}$ are, respectively, the
densities of the ground and excited states of a hydrogen atom, of
ion $H^{+}$, and of free electron in the considered plasma with
given $T$.

\begin{figure}
\centerline{\includegraphics[width=\columnwidth,
height=0.75\columnwidth]{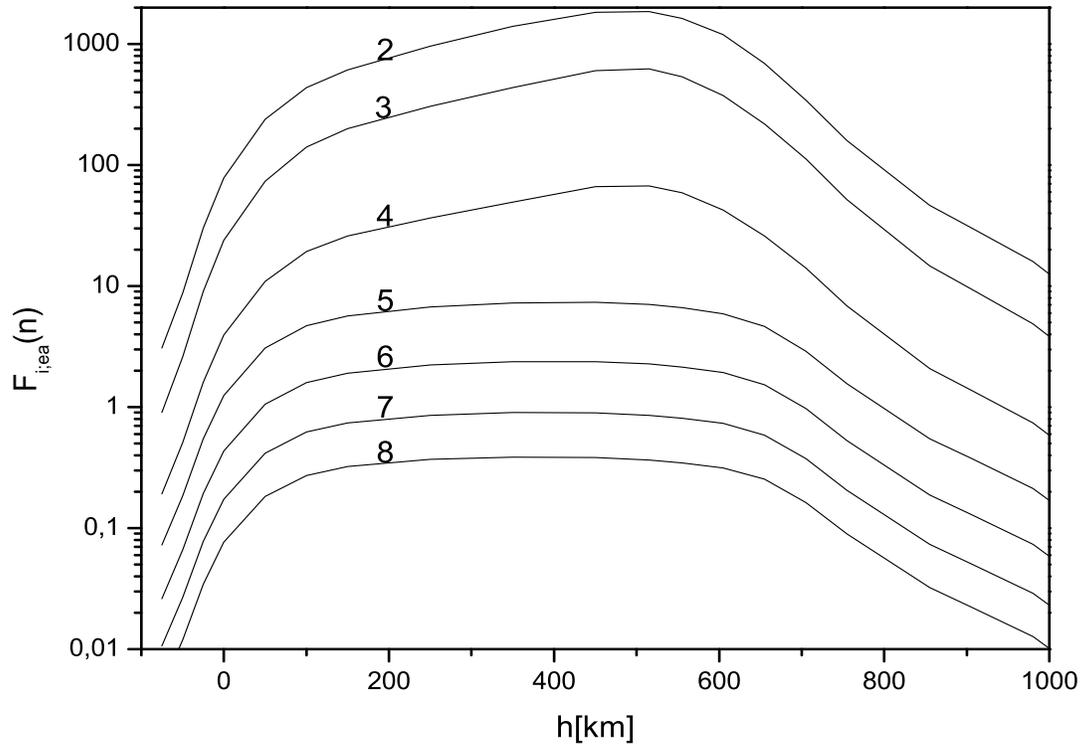}} \caption{Behavior of the
quantity $F_{i;ea}^{(ab)}(n)$ given by Equation (\ref{eq:3i}), as a
function of height h.} \label{fig:Fi_eaN}
\end{figure}

Using these expressions, we will first calculate quantities
$F_{i}(n,T)$ given by
%%%%%%%%%%%%%%%%%%%%%%%%%%%%%%%%%%%%%%%%%%%%%%%%%%%%%%%%%%%%%%%%%%%%%%%%%%
\begin{equation}\label{eq:Fin}
F_{i}(n,T) = \frac{I_{ci}(n,T)}{I_{i;ea}(n,T)} = \frac
{K_{ci}(n,T)}{K_{ea}(n,T)}\cdot {N_{1}}{N_{e}},
\end{equation}
%%%%%%%%%%%%%%%%%%%%%%%%%%%%%%%%%%%%%%%%%%%%%%%%%%%%%%%%%%%%%%%%%%%%%%%%%%
which characterize the relative efficiency of partial
chemi-ionization processes (\ref{eq:1a},\ref{eq:1b}) together and
the impact electron-atom ionization (\ref{eq:3i}) in the considered
plasma. The total chemi-ionization and recombination rate
coefficients $K_{ci}(n,T)$ are determined here the way it is
described in the previous section, and impact ionization rate
coefficients $K_{ea}(n,T)$ are taken from \citet{vri80}. In Figure
\ref{fig:Fi_eaN} the behavior of the quantities $F_{i,ea}(n,T)$ for
$2\le n \le 8$ as functions of height $h$ is shown, according to the
data ($N_{1}$, $N_{e}$ and $T$) from \citet{ver81} for solar
photosphere. One can see that the efficiency of the considered
chemi-ionization processes in comparison with the electron-atom
impact ionization is dominant for 2$\le n \le$6 and becomes
comparable for $n = 7$ and $8$.

\begin{figure}
\centerline{\includegraphics[width=\columnwidth,
height=0.75\columnwidth]{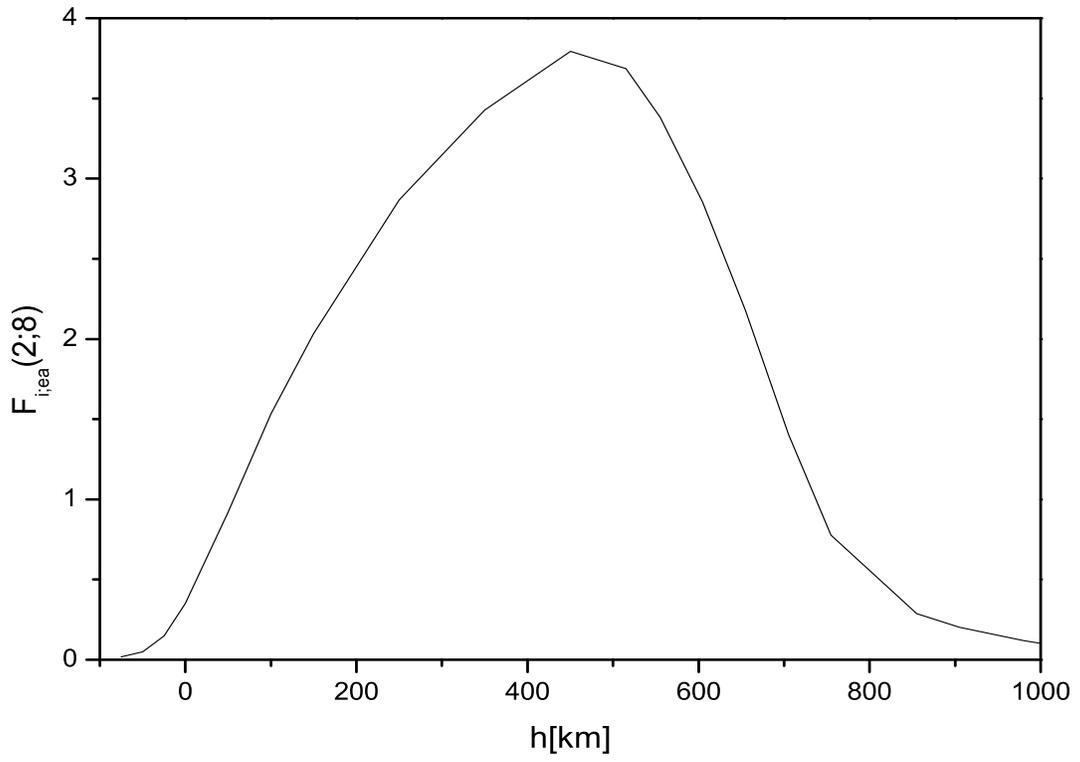}} \caption{Behavior of the
quantity $F_{i;ea}(2;8)$ given by Eq.~(\ref{eq:Fiea2-8}), as a
function of height h} \label{fig:Fi_ea}
\end{figure}

However, in order to compare the relative influence of the
chemi-ionization processes (\ref{eq:1a}) and (\ref{eq:1b}) together
to that of the impact electron-atom ionization process (\ref{eq:3i})
on the whole block of the excited hydrogen atom states with $2\le n
\le 8$, we will calculate  quantity $F_{i,ea;2-8}(T)$, given by
%%%%%%%%%%%%%%%%%%%%%%%%%%%%%%%%%%%%%%%%%%%%%%%%%%%%%%%%%%%%%%%%%%%%%%%%%%
\begin{equation}\label{eq:Fiea2-8}
F_{i,ea;2-8}(T) = \frac{\sum\limits_{n=2}^{8}
I_{ci}(n,T)}{\sum\limits_{n=2}^{8} I_{i;ea}(n,T)} = \frac
{\sum\limits_{n=2}^{8} K_{ci}(n,T)\cdot N_{n}}{\sum\limits_{n=2}^{8}
K_{ea}(n,T)\cdot N_{n}}\cdot {N_{1}}{N_{e}},
\end{equation}
%%%%%%%%%%%%%%%%%%%%%%%%%%%%%%%%%%%%%%%%%%%%%%%%%%%%%%%%%%%%%%%%%%%%%%%%%%
which can reflect the influence of the existing populations of
excited hydrogen atom states within a non-LTE model of solar
atmosphere. In Figure \ref{fig:Fi_ea} the behavior of the quantity
$F_{i,ea;2-8}(T)$ as functions of height $h$ is shown according to
the same data from \citet{ver81}. As one can see, the real influence
of the chemi-ionization processes on the total populations of states
with $2 \le n \le 8$ remains dominant with respect to the concurrent
electron-atom impact ionization processes almost in the whole
photosphere (50 km $\lesssim h \lesssim$ 750 km). This means that
the chemi-ionization processes influence the radiative properties of
the whole solar atmosphere in the optical region considerably.

\begin{figure}
\centerline{\includegraphics[width=\columnwidth,
height=0.75\columnwidth]{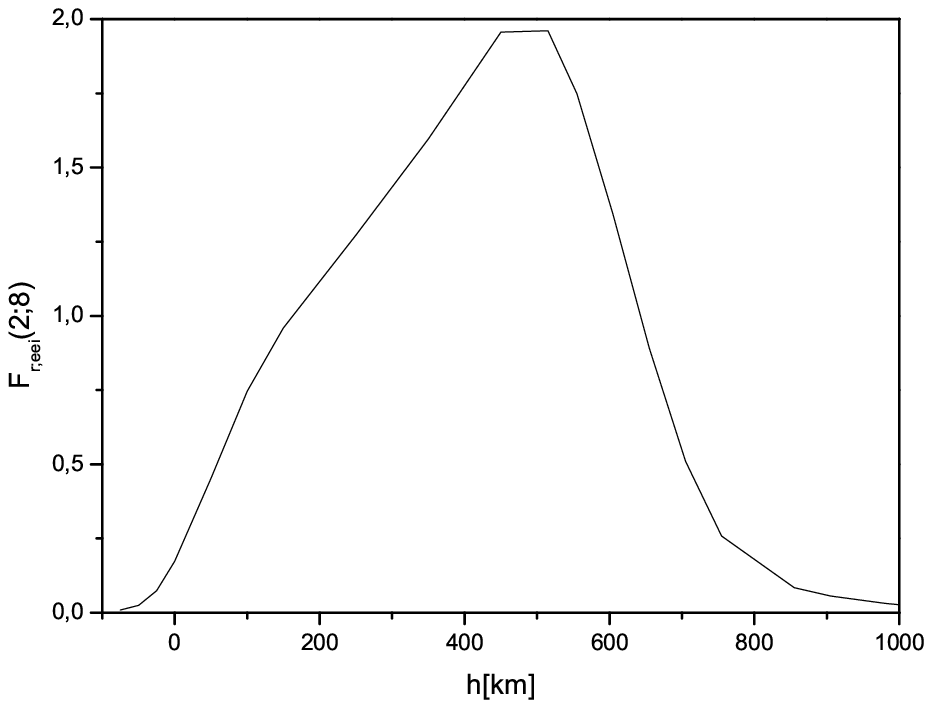}} \caption{Behavior of the
quantity $F_{r;eei}(2;8)$ given by Equation (\ref{eq:Freei2-8}), as
a function of height h.} \label{fig:Fr_eei}
\end{figure}

Then, in order to compare the relative influence of
chemi-recombination processes (\ref{eq:2a}) and (\ref{eq:2b})
together and electron - electron - $H^{+}$ ion recombination process
(\ref{eq:4r}) on the same block of excited hydrogen atom states with
$2\le n \le 8$, we calculated quantity $F_{r,eei;2-8}(T)$, given by
%%%%%%%%%%%%%%%%%%%%%%%%%%%%%%%%%%%%%%%%%%%%%%%%%%%%%%%%%%%%%%%%%%%%%%%%%%
\begin{equation}\label{eq:Freei2-8}
F_{r,eei;2-8}(T) = \frac{\sum\limits_{n=2}^{8}
I_{cr}(n,T)}{\sum\limits_{n=2}^{8} I_{r;eei}(n,T)} = \frac
{\sum\limits_{n=2}^{8} K_{cr}(n,T)}{\sum\limits_{n=2}^{8}
K_{eei}(n,T)}\cdot \frac {N_{1}} {N_{e}},
\end{equation}
%%%%%%%%%%%%%%%%%%%%%%%%%%%%%%%%%%%%%%%%%%%%%%%%%%%%%%%%%%%%%%%%%%%%%%%%%%
taking rate coefficients $K_{eei}(n,T)$ also from \citet{vri80}. In
Figure \ref{fig:Fr_eei} the behavior of this quantity as a function
of height $h$ is shown. One can see that the considered
chemi-recombination processes dominate with respect to the
concurrent electron-electron-ion recombination processes within the
region 100 km $\lesssim h \lesssim$ 650 km. Consequently, the
considered chemi-recombination processes are also very significant
for the optical properties of the solar photosphere.

\begin{figure}
\centerline{\includegraphics[width=\columnwidth,
height=0.75\columnwidth]{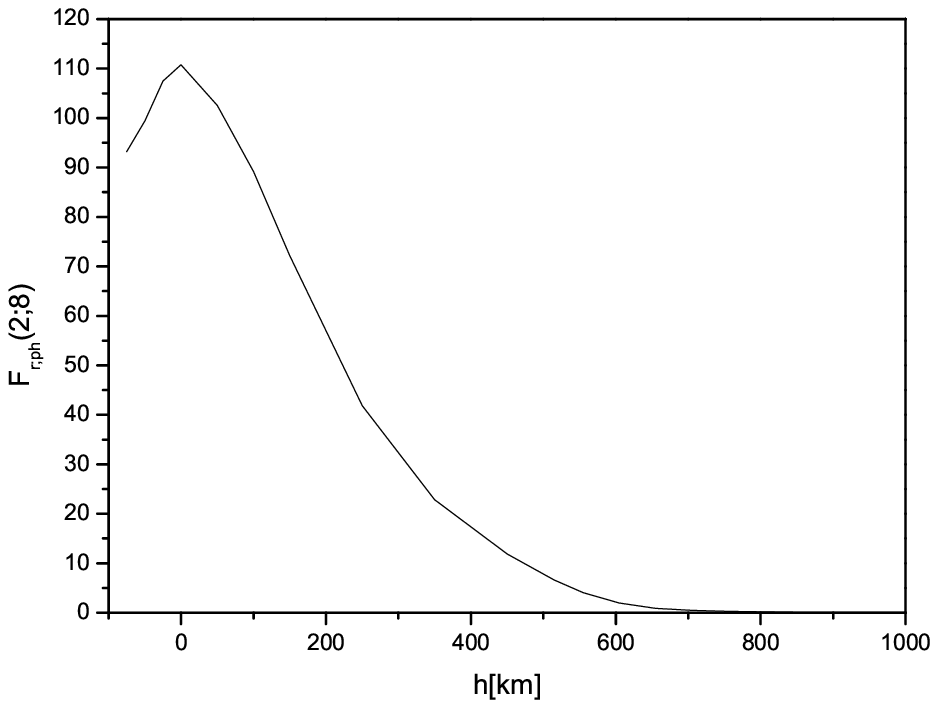}} \caption{Behavior of the
quantity $F_{r;ph}(2;8)$ given by Equation (\ref{eq:Frph2-8}), as a
function of height h.} \label{fig:Fr_ph}
\end{figure}

Finally, we compared the relative influence of chemi-recombination
processes (\ref{eq:2a}) and (\ref{eq:2b}) together and
photo-recombination  electron - $H^{+}$ ion process (\ref{eq:5r}),
also within the block of the excited hydrogen atom states with $2\le
n \le 8$. For that sake we calculated quantity $F_{r,ph;2-8}(T)$,
given by
%%%%%%%%%%%%%%%%%%%%%%%%%%%%%%%%%%%%%%%%%%%%%%%%%%%%%%%%%%%%%%%%%%%%%%%%%%
\begin{equation}\label{eq:Frph2-8}
F_{r,ph;2-8}(T) = \frac{\sum\limits_{n=2}^{8}
I_{cr}(n,T)}{\sum\limits_{n=2}^{8} I_{r;ph}(n,T)} = \frac
{\sum\limits_{n=2}^{8} K_{cr}(n,T)}{\sum\limits_{n=2}^{8}
K_{ph}(n,T)}\cdot N_{1},
\end{equation}
%%%%%%%%%%%%%%%%%%%%%%%%%%%%%%%%%%%%%%%%%%%%%%%%%%%%%%%%%%%%%%%%%%%%%%%%%%
taking rate coefficients $K_{ph}(n,T)$ from \citet{sob79}. This is
necessary since in \citep{mih97} only the states $4 \le n \le 8$
were considered. Still, it was a natural expectation that the
inclusion of states with $n = 2$ and $3$ will increase the influence
of photo-recombination  electron - ion processes. The behavior of
quantity $F_{r,ph;2-8}(T)$ as a function of $h$ is shown in Figure
\ref{fig:Fr_ph}. One can see that here a domination of the
chemi-recombination processes with $2 \le n \le 8$ over the
electron-ion photo-recombination processes is confirmed (although to
a slightly lesser extent) in a significant part of the photosphere
(-50 km $\lesssim h \lesssim$ 600 km).

\section{Conclusion}
The obtained results demonstrate the fact that the considered
chemi-ionization/re\-com\-bi\-na\-ti\-on processes must have a very
significant influence on the optical properties of the solar
photosphere in comparison to the concurrent electron-atom impact
ionization and electron-ion recombination processes. Thus it is
shown that the importance of these processes for non-LTE modeling of
solar atmosphere should be necessarily investigated.

%\bibliographystyle{aa.bst}
%\bibliography{b.bib}

\newcommand{\noopsort}[1]{} \newcommand{\printfirst}[2]{#1}
  \newcommand{\singleletter}[1]{#1} \newcommand{\switchargs}[2]{#2#1}

\acknowledgments

This work was supported by the Ministry of Science and Technological
Development of Serbia as a part of the project "Influence of
collisional processes on astrophysical plasma line shapes" (Project
number 176002).

\clearpage

\begin{table}[htbp]
\begin{center}
\caption{Calculated Values of Coefficient $K_{ci}$[cm$^3$/s] as a
Function of $n$ and $T$} \label{tab:Kci}
\begin{tabular}
{c c c c c c c c} \hline
   & \multicolumn{7}{c}{n}\\
   \cline{2-8}
  T[K]  &     2&         3&         4&         5&         6&         7&         8\\
  \hline
  4000& 0.150E-11& 0.619E-09& 0.126E-08& 0.576E-09& 0.554E-09& 0.463E-09& 0.366E-09\\
  4250& 0.202E-11& 0.549E-09& 0.106E-08& 0.617E-09& 0.583E-09& 0.482E-09& 0.378E-09\\
  4500& 0.260E-11& 0.501E-09& 0.900E-09& 0.656E-09& 0.611E-09& 0.500E-09& 0.389E-09\\
  4750& 0.324E-11& 0.488E-09& 0.833E-09& 0.694E-09& 0.637E-09& 0.517E-09& 0.400E-09\\
  5000& 0.403E-11& 0.495E-09& 0.815E-09& 0.730E-09& 0.662E-09& 0.533E-09& 0.410E-09\\
  5250& 0.504E-11& 0.501E-09& 0.800E-09& 0.765E-09& 0.686E-09& 0.548E-09& 0.420E-09\\
  5500& 0.623E-11& 0.500E-09& 0.782E-09& 0.799E-09& 0.709E-09& 0.563E-09& 0.428E-09\\
  5750& 0.756E-11& 0.493E-09& 0.764E-09& 0.832E-09& 0.731E-09& 0.576E-09& 0.437E-09\\
  6000& 0.909E-11& 0.490E-09& 0.757E-09& 0.864E-09& 0.752E-09& 0.589E-09& 0.445E-09\\
  6250& 0.108E-10& 0.502E-09& 0.766E-09& 0.895E-09& 0.772E-09& 0.602E-09& 0.453E-09\\
  6500& 0.128E-10& 0.519E-09& 0.783E-09& 0.924E-09& 0.791E-09& 0.613E-09& 0.460E-09\\
  7000& 0.175E-10& 0.540E-09& 0.808E-09& 0.981E-09& 0.827E-09& 0.635E-09& 0.473E-09\\
  7500& 0.232E-10& 0.574E-09& 0.848E-09& 0.103E-08& 0.860E-09& 0.655E-09& 0.485E-09\\
  8000& 0.300E-10& 0.609E-09& 0.891E-09& 0.108E-08& 0.892E-09& 0.674E-09& 0.497E-09\\
  8500& 0.380E-10& 0.650E-09& 0.939E-09& 0.113E-08& 0.920E-09& 0.691E-09& 0.507E-09\\
  9000& 0.470E-10& 0.688E-09& 0.986E-09& 0.118E-08& 0.948E-09& 0.707E-09& 0.516E-09\\
  9500& 0.574E-10& 0.733E-09& 0.104E-08& 0.122E-08& 0.973E-09& 0.722E-09& 0.525E-09\\
 10000& 0.689E-10& 0.787E-09& 0.109E-08& 0.126E-08& 0.997E-09& 0.736E-09& 0.533E-09\\
 \hline
 \end{tabular}
 \end{center}
\end{table}

\begin{table}[htbp]
\begin{center}
\caption{Calculated Values of Recombination Coefficient
$K_{cr}$[cm$^6$/s] as a Function of $n$ and $T$} \label{tab:Kcr}
\begin{tabular}
{c c c c c c c c} \hline
   & \multicolumn{7}{c}{n}\\
   \cline{2-8}
  T[K]&     2&         3&         4&         5&         6 &        7&         8    \\
  \hline
  4000& 0.190E-27& 0.732E-27& 0.390E-27& 0.114E-27& 0.977E-28& 0.831E-28& 0.709E-28\\
  4250& 0.130E-27& 0.458E-27& 0.257E-27& 0.102E-27& 0.880E-28& 0.753E-28& 0.645E-28\\
  4500& 0.918E-28& 0.305E-27& 0.177E-27& 0.914E-28& 0.799E-28& 0.688E-28& 0.591E-28\\
  4750& 0.666E-28& 0.223E-27& 0.135E-27& 0.828E-28& 0.730E-28& 0.631E-28& 0.544E-28\\
  5000& 0.506E-28& 0.174E-27& 0.110E-27& 0.755E-28& 0.671E-28& 0.582E-28& 0.503E-28\\
  5250& 0.403E-28& 0.138E-27& 0.912E-28& 0.693E-28& 0.619E-28& 0.540E-28& 0.467E-28\\
  5500& 0.331E-28& 0.111E-27& 0.763E-28& 0.639E-28& 0.575E-28& 0.502E-28& 0.436E-28\\
  5750& 0.275E-28& 0.889E-28& 0.645E-28& 0.592E-28& 0.535E-28& 0.469E-28& 0.407E-28\\
  6000& 0.233E-28& 0.731E-28& 0.558E-28& 0.551E-28& 0.500E-28& 0.440E-28& 0.382E-28\\
  6250& 0.201E-28& 0.627E-28& 0.498E-28& 0.514E-28& 0.469E-28& 0.413E-28& 0.360E-28\\
  6500& 0.176E-28& 0.548E-28& 0.451E-28& 0.482E-28& 0.441E-28& 0.389E-28& 0.339E-28\\
  7000& 0.139E-28& 0.421E-28& 0.374E-28& 0.427E-28& 0.393E-28& 0.348E-28& 0.304E-28\\
  7500& 0.114E-28& 0.341E-28& 0.322E-28& 0.382E-28& 0.354E-28& 0.314E-28& 0.275E-28\\
  8000& 0.964E-29& 0.284E-28& 0.283E-28& 0.345E-28& 0.321E-28& 0.286E-28& 0.250E-28\\
  8500& 0.834E-29& 0.243E-28& 0.253E-28& 0.314E-28& 0.293E-28& 0.261E-28& 0.229E-28\\
  9000& 0.731E-29& 0.211E-28& 0.229E-28& 0.287E-28& 0.269E-28& 0.240E-28& 0.211E-28\\
  9500& 0.654E-29& 0.187E-28& 0.209E-28& 0.264E-28& 0.248E-28& 0.222E-28& 0.195E-28\\
 10000& 0.590E-29& 0.169E-28& 0.194E-28& 0.245E-28& 0.230E-28& 0.206E-28& 0.181E-28\\
 \hline
 \end{tabular}
 \end{center}
\end{table}

\begin{table}[htbp]
\begin{center}
\caption{Calculated Values of Coefficient $X^{(a)} \equiv
K_{ci}^{(a)}/K_{ci} = K_{cr}^{(a)}/K_{cr}$ as a Function of $n$ and
$T$.} \label{tab:X}
\begin{tabular}
{c c c c c c c c} \hline
   & \multicolumn{7}{c}{n}\\
   \cline{2-8}
  T[K]&    2 &    3 &    4 &    5 &    6 &    7 &    8 \\
  \hline
  4000& 0.998& 0.955& 0.877& 0.507& 0.408& 0.335& 0.281\\
  4250& 0.969& 0.934& 0.827& 0.484& 0.388& 0.318& 0.266\\
  4500& 0.924& 0.907& 0.765& 0.463& 0.371& 0.303& 0.254\\
  4750& 0.872& 0.881& 0.709& 0.443& 0.354& 0.289& 0.242\\
  5000& 0.819& 0.857& 0.664& 0.425& 0.339& 0.277& 0.231\\
  5250& 0.769& 0.831& 0.619& 0.408& 0.325& 0.265& 0.221\\
  5500& 0.721& 0.800& 0.568& 0.393& 0.312& 0.254& 0.212\\
  5750& 0.673& 0.764& 0.515& 0.378& 0.300& 0.244& 0.203\\
  6000& 0.627& 0.728& 0.466& 0.364& 0.288& 0.235& 0.196\\
  6250& 0.585& 0.699& 0.430& 0.351& 0.278& 0.226& 0.188\\
  6500& 0.546& 0.672& 0.399& 0.339& 0.268& 0.218& 0.182\\
  7000& 0.474& 0.610& 0.336& 0.317& 0.250& 0.204& 0.169\\
  7500& 0.414& 0.558& 0.289& 0.297& 0.235& 0.190& 0.158\\
  8000& 0.363& 0.510& 0.250& 0.280& 0.221& 0.179& 0.149\\
  8500& 0.321& 0.469& 0.220& 0.264& 0.208& 0.169& 0.141\\
  9000& 0.287& 0.429& 0.193& 0.250& 0.197& 0.160& 0.133\\
  9500& 0.258& 0.398& 0.174& 0.237& 0.187& 0.151& 0.126\\
 10000& 0.234& 0.376& 0.160& 0.225& 0.177& 0.144& 0.120\\
\hline
 \end{tabular}
 \end{center}
\end{table}

\end{document}